\documentclass[preprint,showpacs,preprintnumbers,nofootinbib,amsmath,amssymb]{revtex4}
\usepackage{graphicx}% Include figure files
\usepackage{dcolumn}% Align table columns on decimal points
\usepackage{bm}% bold math
\begin{document}
%%%%%%%%%%%%%%%%%%%%%%%%%%%%%%%%%%%%%%%%%%%%%%%%%
\title{Damping of giant dipole resonance in hot rotating nuclei}
\author{N. Dinh Dang$^{1,2}$}
  \email{dang@riken.jp} 
 \affiliation{1) Theoretical Nuclear Physics Laboratory, RIKEN Nishina Center
for Accelerator-Based Science,
2-1 Hirosawa, Wako City, 351-0198 Saitama, Japan\\
2) Institute for Nuclear Science and Technique, Hanoi, Vietnam}
\date{\today}% It is always \today, today,
             %  but any date may be explicitly specified
%%%%%%%%%%%%%%%%%%%%%%%%%%%%%%%%%%%%%%%%%%%%%%%%%
\begin{abstract}
%%%%%%%%%%%%%%%%%%%%%%%%%%%%%%%%%%%%%%%%%%%%%%%%%
The phonon damping model (PDM) is extended to include the effect of 
angular momentum at finite temperature. The model is applied to the study of damping of giant 
dipole resonance (GDR) in hot and noncollectively rotating spherical nuclei. The numerical results obtained for $^{88}$Mo and $^{106}$Sn show that the GDR width increases with both temperature $T$ and angular momentum $M$. At $T>$ 4 MeV and $M\leq$ 60 $\hbar$ the increase in the GDR width slows down for $^{106}$Sn, whereas at $M\geq$ 80$\hbar$ the GDR widths in both nuclei nearly saturate. By adopting the nuclear shear viscosity extracted from fission data at $T=$ 0,  it is shown that the maximal value of the angular momentum for $^{88}$Mo and $^{106}$Sn should be around 46 and 55 $\hbar$, respectively, so that the universal conjecture for the lower bound of the specific shear viscosity for all fluids is not violated up to $T=$ 5 MeV.
\end{abstract}
% PACS, the Physics and Astronomy
                             % Classification Scheme.
%Use showkeys class option if keyword
                              %display desired
\maketitle
%%%%%%%%%%%%%%%%%%%%%%%%%%%%%%%%%%%%%%%%%%%%%%%%%
\section{Introduction}
The giant electric dipole resonance (GDR) in highly excited nuclei has been a subject of a considerable number of theoretical and experimental studies over the last three decades.  At present, comprehensive experimental data on the dependence of the GDR width on temperature $T$ and angular momentum $J$ have been accumulated for a large number of medium and heavy-mass compound nuclei starting from potassium up to lead isotopes
(See Ref. \cite{Schiller} for the recent compilation). This systematic shows that the full width at half maximum (FWHM) of the GDR increases significantly with $T$ at low and moderate $T$, but seems to saturate at high $T$, above $T\sim$ 4 - 5 MeV. At a given value of $T$, the GDR width  also increases with $J$, but this increase is noticeable only starting from $J>$ 25 $\hbar$ for copper and heavier isotopes. 

A number of theoretical approaches has been proposed to describe the behavior of the GDR width as a function of $T$ and $J$ and two monographies have been published on this subject~\cite{Bortignon,Woude}. Among the theoretical approaches to the study of the hot GDR, the phonon damping model (PDM)~\cite{PDM1,PDM2,PDM3,PDMpair} offers a microscopic mechanism of the GDR damping via coupling of the GDR to non-collective particle-hole ($ph$) transitions, which exist already at $T=$ 0, and particle-particle ($pp$) ones, which appear at $T\neq$ 0 because of the distortion of the Fermi surface. The PDM is able to describe consistently both the increase in the GDR width at low and moderate $T$ as well as its saturation at high $T$ by using the single-particle energies, obtained in the Woods-Saxon potentials for protons and neutrons, and a set of 3 parameters, fixed at $T=$ 0, which are the GDR energy before the coupling, the $ph$ and $pp$ coupling constants~\cite{PDM1,PDM2,PDM3}. By including nonvanishing thermal superfluid pairing, the PDM is also able to explain a nearly
constant value of the GDR width at $T\leq$ 1 MeV, found in the inelastic scattering of $^{17}$O on $^{120}$Sn~\cite{PDMpair}. In a recent development, the PDM was employed to calculate the ratio $\eta/s$ of the shear viscosity $\eta$ to the entropy density $s$ directly from the GDR photoabsorption cross sections in medium and heavy spherical nuclei at $T\neq$ 0~\cite{viscosity}. The results obtained show that this ratio, which is called the specific shear viscosity hereafter, decreases with increasing $T$ to reach (1.3 $-$ 4) KSS at T = 5 MeV. The quantity KSS $\equiv\hbar/(4\pi k_B)$ has been conjectured to be the lower bound for all fluids~\cite{KSS}.

A shortcoming of the PDM is that it does not include so far the dependence on angular momentum.  The aim of the present paper is to remove this deficiency by extending the PDM to incorporate the effect of angular momentum to make the model capable to describe the dependence of the GDR width on both $T$ and $J$.
The present extension will also be applied to examine the credibility of the preliminary analysis of the recent data~\cite{Maj}, according to which the GDR width 
extracted at $T\sim$ 4 MeV and $J=$ 44 $\hbar$ seems to be smaller than that measured at $T\sim$ 3 MeV and $J=$ 41 $\hbar$ in $^{88}$Mo.

The paper is organized as follows. The PDM is extended to include the effect of angular momentum in Sec. \ref{formalism}. The analysis of numerical results is presented 
in Sec. \ref{results}. The article is summarized in the last section, 
where conclusions are drawn.
%%%%%%%%%%%%%%%%%%%%%%%%%%%%%%%%%%%%%%
\section{Angular-momentum effect within the PDM}
\label{formalism}
\subsection{Extention of PDM to finite angular momentum}
The axis of quantization (or the laboratory-frame $z$ axis) of a spherical nucleus can always be chosen to coincide with the body-fixed axis. The latter is aligned with the direction of the total angular momentum within the quantum mechanical uncertainty~\cite{Kammuri,Moretto,SCQRPA}.  As the result,  the total angular momentum is completely determined by its $z$-projection $M$ alone. The rotation of  a spherical nucleus (as of a classical system) about the $z$ axis  is called noncollective or single-particle rotation. On the other hand, even for non-spherical nuclei, especially the axially symmetric ones, at high excitation energies and/or high $T$, the values of angular momentum projection on the symmetry axis are mixed among the levels in the region of high level densities, which worsen the axial symmetry. The melting of shell structure will also eventually drive nuclei to their average spherical shape. The PDM is extended here to include the angular momentum effect for such noncolleticve rotation at finite $T$.

For this purpose, the Hamiltonian of a spherical system, which is rotating
about the symmetry $z$-axis, can be written in the following form
\begin{equation}
H=H_{0} - \gamma\hat{M}~,
\label{H0}
\end{equation}
where $H_0$ is the PDM Hamiltonian described in Ref. \cite{PDM1,PDM2}, and $\hat{M}$ represents the total angular momentum $\hat{J}$, which, in this case, coincides with its $z$-projection $M$. The latter is defined as 
\begin{equation}
\hat{M} =\sum_{k>0}m_{k}(N_{k}-N_{-k})~,
\label{M}
\end{equation}
where, for simplicity, the subscripts $k$ 
are used to denote the single-particle states 
$|k,m_{k}\rangle$ in the deformed basis with the angular momentum $k$ and the positive single-particle spin
projection $m_{k}$, whereas the subscripts $-k$  denote the
time-reversal states $|k,-m_{k}\rangle$ ($m_{k}>$ 0). In the spherical basis
this corresponds to $|j,m\rangle$. The deformed basis is preferable here because
the rotation about the $z$-axis eventually resolves the $(2j+1)$ degeneracy of each spherical orbital $j$. The particle number operator $\hat{N}$ consists of
\begin{equation}
\hat{N} = \sum_{k>0}(N_{k}+N_{-k})~,\hspace{5mm} N_{\pm k}=a_{\pm k}^{\dagger}a_{\pm k}~,
\label{N}
\end{equation}
where $a_{\pm k}^{\dagger}$
($a_{\pm k}$) denotes the creation (annihilation) operator of a particle with angular momentum $k$, spin projection $\pm m_k$, and energy $\epsilon_{k}$. 

By using Eqs. (\ref{M}) and (\ref{N}), the Hamiltonian (\ref{H0}) transforms in the deformed basis into
\[
H = \sum_{k>0}(\epsilon_k-\lambda
-\gamma m_k)N_k+\sum_{k>0}(\epsilon_k-\lambda+\gamma m_k)N_{-k}+
\sum_{q}\omega_{q}Q_{q}^{\dagger}Q_{q}
\]
\begin{equation}
+\sum_{k,k'>0}\sum_q{\cal F}_{kk'}^{(q)}(a_{k}^{\dagger}a_{k'}+a_{-k}^{\dagger}a_{-k'})(Q_{q}^{\dagger}+Q_{q})~,
\label{H}
\end{equation}
where $\lambda$ denotes the chemical potential.
The particle ($p$) states are defined as those with $\epsilon_k>\lambda$, whereas the hole ($h$) states are those with $\epsilon_k<\lambda$. The operator $Q_{q}^{\dagger}$ ($Q_{q}$) creates (annihilates) 
a phonon with energy $\omega_{q}$, which describes the collective vibration. Therefore, the right-hand side of Eq. (\ref{H}) describes two mean fields, one of the single particles as the first two terms,  another of the phonon field
associated with the GDR as the third term, and the coupling between them in the last term with matrix elements ${\cal F}_{kk'}^{(q)}$. This coupling causes the damping of the GDR. The effects of angular momentum, which are incorporated in the first two terms, resolve the degeneracy of spherical orbitals. As the result, each of spherical orbital $j$ with energy $\epsilon_j$ splits into $2\Omega_j=2j + 1$ distinctive levels, half of which consists of levels with energies $\epsilon_k+\gamma m_k$ , whereas the other half consists of levels with energies $\epsilon_k-\gamma m_k$, with $k = 1, ..., \Omega/2$, where $\Omega=2\sum_j\Omega_j$ is the total number of levels. Thermal superfluid 
pairing plays the role of keeping the GDR width essentially unchanged at $T\leq$ 1 MeV at $M=$ 0~\cite{PDMpair}. Pairing has a negligible (or no) effect on the GDR width in the region of moderate (high) $T$ and $M$, which is the main focus of the present paper. Therefore it is not included in the calculation of the GDR width here for simplicity. 

The chemical potential $\lambda$ and the rotation frequency $\gamma$ are two Lagrangian multipliers, which are defined from the equations for conservation of angular momentum $M$~\cite{Kammuri, Moretto, SCQRPA} and particle number $N$ as
\begin{equation}
M = \sum_k m_k(f_{k}^{+}-f_{k}^{-})~,\hspace{5mm} N =\sum_k(f_{k}^{+}+f_{k}^{-})~,
\label{M&N}
\end{equation}
where $M=\langle\hat{M}\rangle$, $N=\langle\hat{N}\rangle$, $f_{k}^{\pm} =
\langle N_{\pm k}\rangle$ with $\langle...\rangle\equiv{\rm Tr}[...{\exp}(-\beta H)]/{\rm Tr}[{\exp}(-\beta H)]$ denoting the grand canonical ensemble average with $\beta = T^{-1}$. The PDM calculations at $M=$ 0 showed that, although the particle-phonon coupling in the last term of Eq. (\ref{H}) is sufficient to cause the GDR width of around 4 to 5 MeV for medium and heavy nuclei at $T=$ 0, it is not sufficiently large to cause a noticeable damping of each single-particle state. Therefore, the single-particle occupation numbers $f_k^{\pm}$ can be well approximated with the Fermi-Dirac distribution for nonineracting fermions, namely 
\begin{equation}
f_k^{\pm} =\frac{1}{\exp(\beta E_k^{\mp})+1}~,\hspace{5mm} E_k^{\mp} = \epsilon_k-\lambda\mp\gamma m_k~.
\label{nk}
\end{equation}
By solving Eqs. (\ref{M&N}) at each value of temperature $T$, the chemical potential $\lambda$ and rotation frequency $\gamma$ are defined as functions of $T$. Following the same procedure as that employed to obtain the PDM equations at $M=$ 0, one
now proceeds to derive the strength function and the width of the GDR at 
$M\neq$ 0 below.

The following double-time retarded Green functions are introduced, which 
describe~\cite{PDM1,PDM2} 

1) {\it The propagation of the free phonon}:
\begin{equation}
G_{q}(t-t')=\langle\langle 
Q_{q}(t);Q_{q}^{\dagger}(t')\rangle\rangle~,
\label{Gq}
\end{equation}

2) {\it The transition between a nucleon pair with positive spin $m_k$ and the phonon}:
\begin{equation}
{\cal G}_{k k'q}^{+}(t-t')=\langle\langle 
a_{k}^{\dagger}(t)a_{k'}(t);Q_{q}^{\dagger}(t')\rangle\rangle~,
\label{G+}
\end{equation}

3) {\it The transition between a nucleon pair with negative spin $-m_k$ and the phonon}:
\begin{equation}
{\cal G}_{kk'q}^{-}(t-t')=\langle\langle 
a_{-k}^{\dagger}(t)a_{-k'}(t);Q_{q}^{\dagger}(t')\rangle\rangle~,
\label{G-}
\end{equation}
where the standard notation $G(t,t')\equiv\langle\langle A(t); B(t')\rangle\rangle=-i\theta(t-t')\langle[A(t),B(t')]_{\pm}\rangle$ is used for the double-time retarded Green function 
with $[A(t), B(t')]_{\pm}= A(t)B(t')-B(t')A(t)$ for boson operators, and $A(t)B(t')+B(t')A(t)$ for fermion ones~\cite{Zubarev}. 
Applying the standard procedure of deriving the equation of motion for the double-time Green function  in the form
$i{dG}/{dt} =\delta(t-t')\langle[A(t), B(t)]_{\pm}\rangle +\langle\langle[A(t),H(t)];B(t')\rangle\rangle$ to the Green functions (\ref{Gq}) -- (\ref{G-}) and Hamiltonian (\ref{H}), one obtains an hierarchy of equations. This infinite series contains, beside the Green functions (\ref{Gq}) -- (\ref{G-}), also the higher-order ones. To close this set, a simple decoupling is used by pairing off, wherever possible, the operators referring to the same time, e.g. 
$\langle\langle a^{\dagger}_{\pm k}(t)a_{\pm k'}(t)Q_q(t);Q^{\dagger}(t')\rangle\rangle=\delta_{kk'}f_k^{\pm}G_q(t-t')$. The whole procedure is the same as has been described thoroughly in Refs. \cite{PDM1,PDM2} so it is not repeated here.   The final set contains only 3 coupled equations for three double-time Green functions (\ref{Gq}) -- (\ref{G-}), whose Fourier transforms into the energy variable $E$ are
\begin{equation}
(E-\omega_q)G_q(E) = \frac{1}{2\pi}+\sum_{kk'}{\cal F}^{(q)}_{kk'}[{\cal G}_{kk'q}^{+}(E) +{\cal G}_{kk'q}^{-}(E)]~,
\label{GE}
\end{equation}
\begin{equation}
(E-E_k^{-}+E_{k'}^{-}){\cal G}_{kk'q}^{+}(E)={\cal F}^{(q)}_{kk'}(f^{+}_{k'}-f^{+}_{k})G_q(E)~,
\label{G+E}
\end{equation}
\begin{equation}
(E-E_k^{+}+E_{k'}^{+}){\cal G}_{kk'q}^{-}(E)={\cal F}^{(q)}_{kk'}(f^{-}_{k'}-f^{-}_{k})G_q(E)~.
\label{G-E}
\end{equation}
Expressing ${\cal G}_{kk'q}^{\pm}(E)$  in terms of $G_q(E)$ by using the last two equations, and inserting the results into Eq. (\ref{GE}), one obtain
the final equation for the Green function $G_q(E)$, which describes the phonon propagation, as
\begin{equation}
G_q(E) = \frac{1}{2\pi}\frac{1}{E-\tilde\omega_q}~,\hspace{2mm} \tilde\omega=\omega_q+P_q(E)~,
\hspace{2mm} P_q(E) = \sum_{kk'}[{\cal F}_{kk'}^{(q)}]^{2}\bigg[\frac{f^{+}_{k'}-f^{+}_{k}}
{E-E_k^{-}+E_{k'}^{-}}+\frac{f^{-}_{k'}-f^{-}_{k}}
{E-E_k^{+}+E_{k'}^{+}} \bigg]~.
\label{Gfinal}
\end{equation}
The principal value of the polarization operator $P_q(\omega)$ at a real $\omega$ defines the energy shift from the unperturbed phonon energy $\omega_q$ to $\tilde\omega_q$ under the effect of particle-phonon coupling in the last term of the Hamiltoanian (\ref{H}), whereas the phonon damping $\gamma_q(\omega)$ is defined as the imaginary part of the analytic continuation of $P_q(E)$ into the complex energy plan $E=\omega\pm  i\varepsilon$, that is 
$\gamma_q(\omega) = \Im m P_q(\omega\pm i\varepsilon)$. The final results reads
\begin{equation}
\gamma_q(\omega) = \pi\sum_{kk'}[{\cal F}_{kk'}^{(q)}]^{2}[(f_{k'}^{+}-f_{k}^{+})
\delta(\omega-E_k^{-} + E_{k'}^{-})+(f_{k'}^{-}-f_{k}^{-})
\delta(\omega-E_k^{+} + E_{k'}^{+})]~,
\label{gamma}
\end{equation}
which, by using the $\delta$-function representation $\delta(x) =\lim_{\varepsilon\rightarrow 0}\varepsilon/[\pi(x^{2}+\varepsilon^2)]$, transforms into
\begin{equation}
\gamma_q(\omega) = \varepsilon\sum_{kk'}[{\cal F}_{kk'}^{(q)}]^{2}
\bigg[\frac{f_{k'}^{+}-f_{k}^{+}}
{(\omega-E_k^{-} + E_{k'}^{-})^2+\varepsilon^2}+\frac{f_{k'}^{-}-f_{k}^{-}}{(\omega-E_k^{+} + E_{k'}^{+})^{2}+\varepsilon^{2}}\bigg]~.
\label{gamma1}
\end{equation}

The spectral intensity $J_q(\omega)$ is found by definition from the analytic properties of the Green function $G_q(E)$ (\ref{Gfinal}) as 
\begin{equation}
J_q(\omega) = i\frac{G_q(\omega+i\varepsilon) - G_q(\omega-i\varepsilon)}{{\rm e}^{\beta\omega}-1}=
\frac{1}{\pi}\frac{\gamma_q(\omega)({\rm e}^{\beta\omega}-1)^{-1}}{(\omega-\tilde\omega_q)^2+\gamma_q^2(\omega)}~.
\label{Jq}
\end{equation}
The GDR strength function $S(\omega)$ is obtained from the spectral intensity $J_q(\omega)$
as $S(\omega) = \bar{J}_q(\omega)[\exp(\beta\omega)-1]$, where $\bar{J}_q(\omega)$ 
denotes $J_q(\omega)$ calculated at  the GDR energy $\tilde\omega = E_{GDR}(T)$ at temperature $T$~\cite{PDM1,PDM2,PDM3}.
The result reads
\begin{equation}
S(\omega) =
\frac{1}{\pi}\frac{\gamma_q(\omega)}{[\omega-E_{GDR}(T)]^2+\gamma_q^2(\omega)}~.
\label{SE}
\end{equation}
The FWHM width $\Gamma(T)$ of the GDR 
is defined as a function of $T$ at each value $M$ of the total angular momentum as~\cite{PDM1,PDM2}
\begin{equation}
\Gamma(T) = 2\gamma_q[\omega=E_{GDR}(T)]~.
\label{FWHM}
\end{equation}
%%%%%%%%%%%%%%%%% revision 1 %%%%%%%%%%%%%%%%%%%%%%%%%%%%%%%%%%%%
This width consists of the quantal and thermal parts [See Eqs. (1a) - (1c) of Ref. \cite{PDMpair}].
The quantal part is the spreading width $\Gamma^{\downarrow}$, caused by coupling of the GDR to $ph$ configurations.  The thermal part comes from coupling of the GDR to $pp$ and $hh$ configurations, which appear due to the distortion of the Fermi surface at $T\neq$ 0. The escape width $\Gamma^{\uparrow}$, which arises due to coupling to continuum and is related to the direct decay by particle emission,  is usually small (in the order of few hundreds keVs)~\cite{Giai}, and there is no evidence that it is sensitive to the change of $T$ in medium and heavy nuclei. In the numerical calculations within the PDM, the effect of the escape width $\Gamma^{\uparrow}$ is taken into account via the smoothing parameter $\varepsilon$ in Eq. (\ref{gamma1}), which usually does not exceed 1 MeV (See Sec. \ref{results} B).

%%%%%%%%%%%%%%%%%%%%%%%%%%% revision2 %%%%%%%%%%%%%%%%%%%%%%%%%%%%%
Finally, it is worth mentioning that Ref. \cite{Chomaz} proposed an evaporation width $\Gamma_{ev}$ of the compound nucleus states to be added twice to the total GDR width (See also p. 226 of Ref. \cite{Bortignon}). This evaporation width comes from the quantum mechanical uncertainty principle, according to which, the energies of the compound nucleus with a finite lifetime $t$ cannot be known with an accuracy better than $\Gamma_{ev}\sim\hbar/t$. Once this width is incorporated in the calculations of the transition probabilities that define the amplitudes of the GDR photoabsorption cross section, it gives the natural width of 2$\Gamma_{ev}$, which is approximately the sum of the widths of the initial and final states.  In heavy nuclei at high excitation energies, the width 2$\Gamma_{ev}$ can amount to few MeVs (See, e.g., Tabs. 3 and 4 as well as Figs. 5 and 8 of Ref. \cite{Ormand}).  The PDM does not takes the evaporation width $\Gamma_{ev}$ into account. However, it should be noticed that,  a complete procedure requires the comparison of not just the GDR energy and width, but of the entire GDR shape generated by the theoretical strength functions with that obtained from the measured $\gamma$-ray spectra.  For this purpose, the authors of Ref. \cite{Gervais} proposed a method, which incorporates the theoretical strength functions directly into all the decay steps of the full statistical calculations. 
This method allows to test the contribution of the evaporation width as well. As a matter of fact, by using the evaporation width 2$\Gamma_{ev}$ obtained in Refs. \cite{Chomaz,Ormand}, the authors of this method found that the contribution to the total $\gamma$-spectrum by the evaporation with is small relatively to the total spectrum including all the decay steps. The overall high energy $\gamma$-ray spectra resulting from the complete CASCADE calculations including the evaporation width 2$\Gamma_{ev}$ turn out to be essentially identical to those obtained by neglecting this evaporation width even up to the excitation energy  higher than 120 MeV for $^{120}$Sn (i.e. at $T>$ 3.3 MeV). In Ref. \cite{Eisenman}, by using the same method,  the PDM strength functions were also incorporated into all the steps of the CASCADE calculations, and a reasonable agreement between theory and experiment were found,  especially at $T\geq$ 2 MeV. These results indicate that the effect of the evaporation width 2$\Gamma_{ev}$ may become substantial only at much  higher values of temperature and angular momentum ($T\gg$ 3.3 MeV and $J\gg$ 30$\hbar$). Further experimenal tests of this effect at high excitation energies are still required.

%%%%%%%%%%%%%%%%%%%%%%%%%%%%%%%%%%%%%%%%%%%%%%%%%%%%%%%%%%%%%%%%%%%%%%%%%%
\subsection{Specific shear viscosity}
Knowing the GDR parameters $E_{GDR}(T)$ and $\Gamma(T)$, one can calculate the GDR photoabsorption cross-section $\sigma_{GDR}(\omega,T)$, and the shear viscosity $\eta(T)$ as functions of $T$ at each $M$ by using the Green-Kubo formula in combination with the  fluctuation-dissipation theorem (See Ref. \cite{viscosity} for details). The result yields [See Eq. (6) of Ref. \cite{viscosity}] 
\begin{equation}
\eta(T)=\lim_{\omega\rightarrow 0}\frac{\sigma_{GDR}(\omega,T)}{C} = \eta(0)\frac{\Gamma(T)}{\Gamma(0)}
\frac{E_{GDR}(0)^{2}+[\Gamma(0)/2]^{2}}{E_{GDR}(T)^{2}+[\Gamma(T)/2]^{2}}~,
\label{eta}
\end{equation}
where the normalization parameter $C$ is chosen to reproduce the value $\eta(0)$ of $\eta(T)$ at $T=$ 0, that is
$C=\lim_{\omega\rightarrow 0}[\sigma_{GDR}(\omega,T=0)]/\eta(0)$. 

To obtain the specific shear viscosity $\eta/s$ one needs to know the entropy density (entropy per volume $V$) $s = {\cal S}/{V} = \rho{\cal S}/{A}$ with the nuclear density $\rho=$ 0.16 fm$^{-3}$ and nuclear mass number $A$.  The entropy ${\cal S}$ at 
temperature $T$ is obtained by integrating the Clausius definition of entropy as 
${\cal S} = \int_{0}^{T}d\tau[{\tau}^{-1}\partial{\cal E}/{\partial\tau}]$,
where ${\cal E}$ is the total energy of the system at temperature 
$\tau$. By using this definition and taking the grand-canonical-ensemble average of the PDM Hamiltonian (\ref{H}), it follows that 
${\cal S}={\cal S}_{\alpha}+{\cal S}_{Q}$, where $S_{\alpha}$ and $S_{Q}$ are the
entropies of the quasiparticle and phonon mean fields, respectively.
The quasiparticle entropy ${\cal S}_{\alpha}$ is given in units of Boltzmann
constant $k_{B}$ as
\begin{equation}
{\cal S}_{\alpha}=-\sum_k[n_k^+ {\rm ln} n_k^+ + (1-n_k^+){\rm ln}(1-n_k^+)+ n_k^{-} {\rm ln} n_k^{-} + (1-n_k^{-}){\rm ln}(1-n_k^{-})]
    \label{S}
    \end{equation}
    where $n_k^{\pm} =\{\exp[\beta (E_k\mp\gamma m_k)+1\}^{-1}$ are the quasiparticle occupation numbers on the $k$-th level with quasiparticle energies $E_k=\sqrt{(\epsilon_k-\lambda)^2+\Delta(T)}$. The pairing gap $\Delta(T)$ is needed in the calculation of the entropy in open-shell nuclei to ensure that ${\cal S}_{\alpha}$ goes smoothly to zero at $T\rightarrow$ 0. At $T>T_c\simeq 0.57\Delta(0)$, where the gap vanishes as in the BCS theory (or remains finite but small as in the case when thermal fluctuations of the pairing field are included), $n_k^{\pm}$ coincide with (or approach to) $f_k^{\pm}$ and Eq. (\ref{S}) 
yields the single-particle entropy. Because the effect of thermal fluctuations on the entropy is not large at moderate and high $T$, the present paper does not take them into account, adopting the pairing gap $\Delta(T)$ obtained from the BCS theory including the angular momentum effect at $T\neq$ 0~\cite{Kammuri,Moretto}. 

The phonon entropy ${\cal S}_{Q}$ is given as ${\cal S}_Q=\sum_q [(1+\nu_q)\ln(1+\nu_q)-\nu_q\ln n_q]$, where $\nu_q\simeq[\exp(\beta\omega_q)-1]^{-1}$ is the phonon occupation number approximated with the Bose-Einstein distribution for noninteracting bosons. However, in practical calculations of GDR within the PDM, $\omega_q$ is close to $E_{GDR}(T)\gg T$, which means that $\nu_q$ are small~\cite{viscosity}. As the result, ${\cal S}_Q$ for GDR is negligible as compared to ${\cal S}_{\alpha}$, allowing ${\cal S}_Q$ to be safely omitted to give ${\cal S}\simeq {\cal S}_{\alpha}$ in the present calculations.  
%%%%%%%%%%%%%%%%%%%%%%%%%%%%%%%%%%%%%%%%
\section{Analysis of numerical results}
\label{results}
\subsection{Ingredients of numerical calculations}
%%%%%%%%%%%%%%%%%%%%%%%%%%%%%%%%%%%%%%
The single-particle energies $\epsilon_{j}$ are obtained from the spherical Woods-Saxon potentials for neutrons and protons.  At $M=$ 0, these $(2j+1)$-degenerate levels span a large space starting from the bottom $1s_{1/2}$ level located at around $-40$ MeV for neutrons and $-30$ MeV for protons up to around 22 MeV for $^{88}$Mo and around 26 MeV (18 MeV) for protons (neutrons) in $^{106}$Sn. They are kept unchanged with $T$ based on the estimation within the temperature-dependent self-consistent Hartree-Fock calculations~\cite{Quentin}, which have demonstrated that the single-particle energies in heavy nuclei weakly change with $T$ up to $T\sim$ 5 MeV.

The PDM assumes that the matrix elements ${\cal F}_{ph}^{(q)}$ for the coupling of the GDR to non-collective $ph$ configurations, causing the quantal width already at $T=$ 0, are all equal to parameter ${\cal F}_1$,  whereas those for the coupling of the GDR to $pp$ ($hh$) configurations, ${\cal F}_{pp}^{(q)}$ and ${\cal F}_{hh}^{(q)}$, causing the thermal width at $T\neq$ 0, are all equal to parameter ${\cal F}_2$. The justification of such assumption has been discussed in detail previously so it is not repeated here [See, e.g., Sec. II B of Ref. \cite{PDMpair} for the details]. With this assumption the PDM has three parameters, ${\cal F}_1$, ${\cal F}_2$, and the unperturbed phonon energy $\omega_{q}$. The latter, in the case of GDR ($q=1$), is chosen to be close to the $E_{GDR}(0)$. The parameters ${\cal F}_1$ and ${\cal F}_2$ are adjusted to reproduce the experimental value of the GDR width at ($T=$ 0, $M=$ 0), and so that the GDR energy $E_{GDR}(T)$ does not change appreciably with $T$.  
At $M=$ 0, $E_{GDR}(T)$ is usually determined from the equation $E_{GDR}(T) = \omega_q + P_q[E_{GDR}(T)]$. At $M\neq$ 0, to avoid solving  this equation at each value of $M$, it is assumed, based on experimental systematics, that $E_{GDR}$ does not depend on $T$ and $M$. Therefore, the values of parameters ${\cal F}_1$ and ${\cal F}_2$, chosen at $M=$ 0, are adopted, whereas the experimental value of $E_{GDR}(T)=E_{GDR}(0)$, which is 15 MeV for $^{88}$Mo and 15.5 MeV for $^{120}$Sn, is used in Eq. (\ref{SE}) as $T$ and $M$ vary. 

For the calculations of the entropies in $^{88}$Mo and $^{106}$Sn, the pairing interaction parameters are chosen to reproduce the experimental values of the neutron and/or proton gaps at $T=$ 0 and $M=$ 0. Hence, the present paper adopts $G_Z=$ 0.154 MeV for protons and $G_N=$ 0.153 MeV for neutrons in $^{88}$Mo to obtain the pairing gaps
 $\Delta_Z(0)\simeq$ 1.25 MeV and $\Delta_N(0)\simeq$ 1.39 MeV at $M=$ 0. For the proton closed-shell $^{106}$Sn the value $G_N=$ 0.124 MeV is used to give $\Delta_N(0)\simeq$ 1.25 MeV. These values of pairing gaps agree with the empirical data~\cite{Bohr}. As for $\eta(0)$ in Eq. (\ref{eta}), the value $\eta(0) = 1u$, extracted from the GDR data at $T=$ 0~\cite{Auerbach1}, and the lowest value $\eta(0)=0.6u$, obtained by fitting the fission data with $u = 10^{-23}$ Mev s
fm$^{-3}$ are used [See Sec. II A of Ref. \cite{viscosity} for the detail discussion on the selection of $\eta(0)$].
%%%%%%%%%%%%%%%%%%%%%%%%%%%%%%%%%%%%
\subsection{GDR Strength function and width as functions of $T$ and $M$}
\begin{figure}
     \includegraphics[width=16.0 cm]{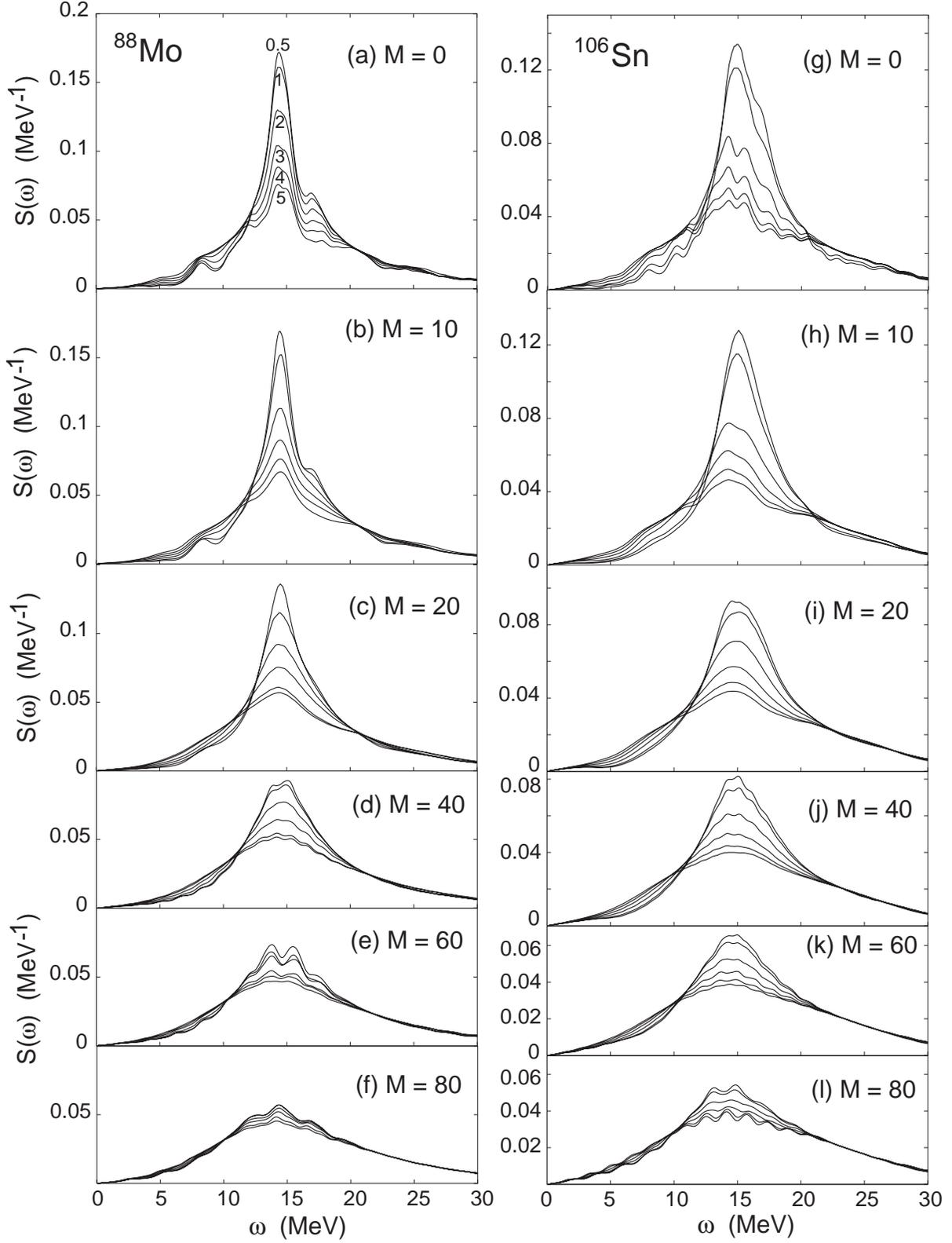}
     \caption{(Color online) GDR strength functions  $S(\omega)$ 
     for $^{88}$Mo [(a) -- (f)] and $^{106}$Sn [(g) -- (l)] at $T=$ 0.5, 1, 2, 3, 4, and 5 MeV as shown at the curves in (a), and $M=$ 0, 10, 20, 40, 60, and 80 $\hbar$ as shown in the panels.}    
      \label{stre}
\end{figure}

\begin{figure}
     \includegraphics[width=16.0 cm]{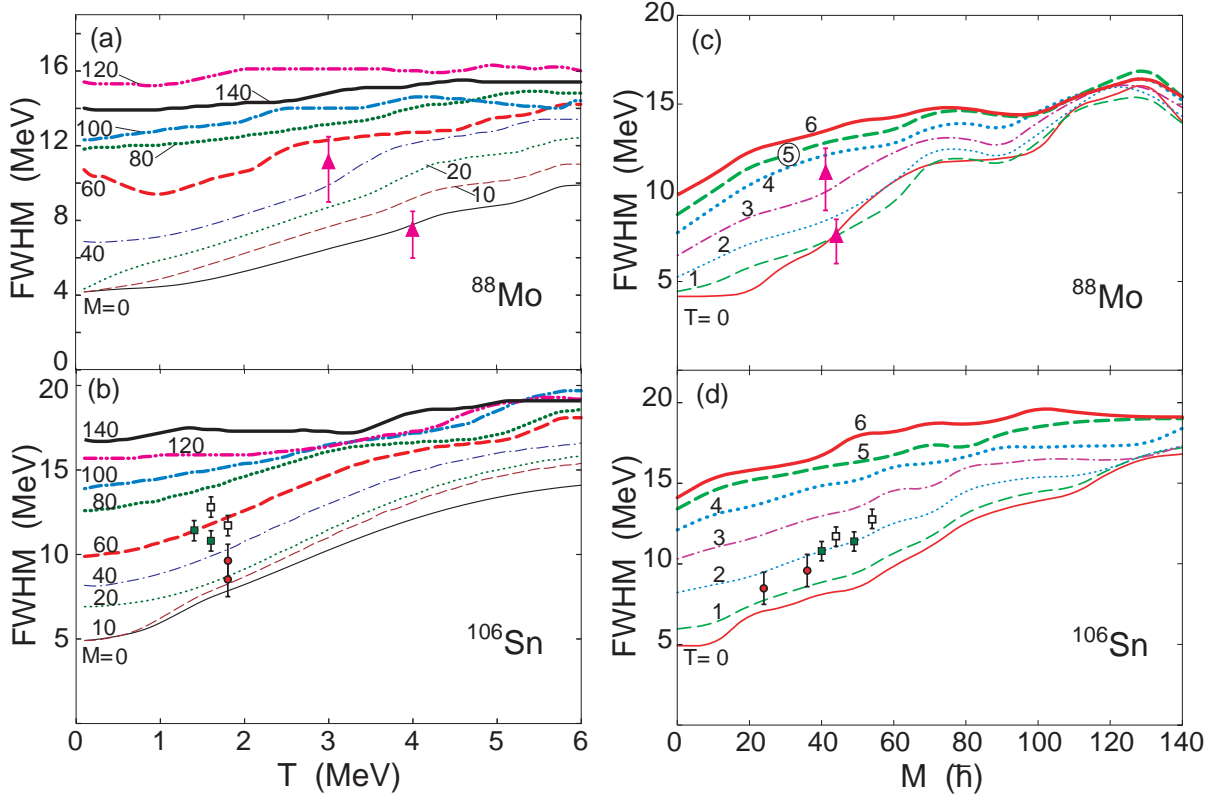}
     \caption{(Color online) FWHM of GDR for $^{88}$Mo [(a), (c)] and $^{106}$Sn [(b), (d)] as a function of $T$
     at several values of $M$ (in $\hbar$) shown at the curves [(a) and (b)], and as a function of $M$ at several values of $T$ (in MeV) shown at the curves [(c) and (d)]. The experimental data for GDR in $^{88}$Mo (triangles), $^{106}$Sn (solid circles), $^{109,110}$Sn (solid and open boxes) are adapted from Ref. \cite{Maj, Mattiuzzi,Bra}, respectively.}    
      \label{width}
\end{figure}

%%%%%%%%%%%%%%%%%%%%%%%%%%%%%%%%%%%%%
Shown in  Figs. \ref{stre} and \ref{width} are the strength functions $S(\omega)$, obtained from Eq. (\ref{SE}), and the width, obtained from Eq. (\ref{FWHM}), of the GDR in
$^{88}$Mo and $^{106}$Sn at various values of temperature $T$ and angular momentum $M$. A smoothing parameter $\varepsilon =$ 0.5 MeV in Eq. (\ref{gamma1}) was used in calculations. The results of the GDR width (\ref{FWHM}) do not change significantly within the interval $0.1\leq \varepsilon\leq$ 1 MeV, but the strength function $S(\omega)$ is smoother by using a larger $\varepsilon$. 


\begin{figure}
     \includegraphics[width=16.0 cm]{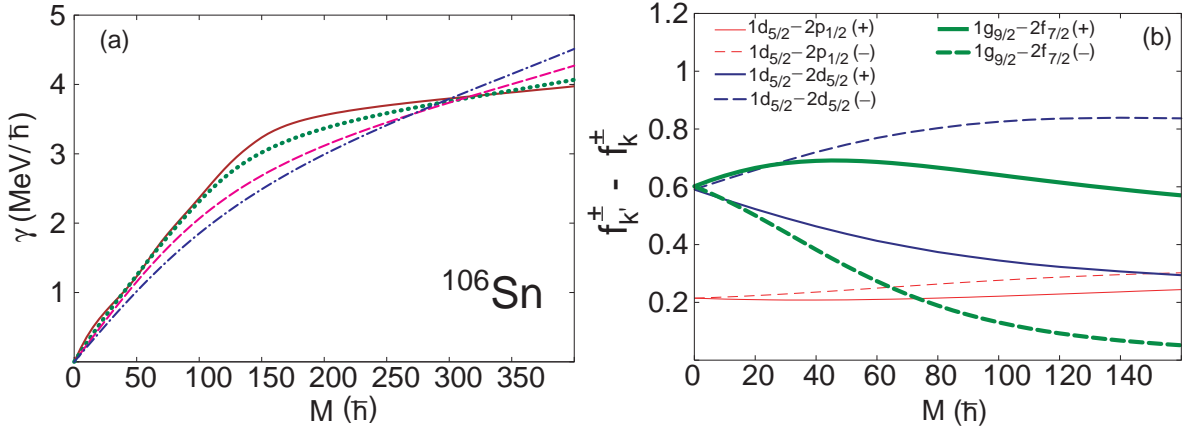}
     \caption{(Color online) (a) Rotation frequency $\gamma$ as a function of angular momentum $M$ at several temperatures $T=$ 0.4 (red solid), 1 (green dotted), 2 (purple dashed), and 4 (blue dot-dashed) MeV in $^{106}$Sn. (b) Differences  $f_{k'}^{\pm}-f_{k}^{\pm}$ for the pair of neutron levels $(1d_{5/2}-2p_{1/2})$, $(1d_{5/2}-2d_{5/2})$, and $(1g_{9/2}-2f_{7/2})$ in $^{106}$Sn,  calculated at $T=$ 4 MeV, as functions of $M$. The signs (+) and (-) at the notations for the lines correspond to
     $(f_{k'}^{+}-f_k^{+})$ and $(f_{k'}^{-}-f_k^{-})$, respectively.}    
      \label{ff}
\end{figure}

Three features are clearly seen from these figures. The first one is that the GDR width increases with both $T$ and $M$. This increase is stronger at low $T$ and $M$. The second feature is that, while the GDR shape becomes smoother  as $T$ and $M$ increase, the smoothing caused by the angular momentum seems to be stronger than that caused by thermal effects (Fig. \ref{stre}). The third feature is that the GDR width approaches a saturation at moderate and high $T$ or $M$. As a function of $T$, the width saturation begins at $T>$ 4 MeV in $^{106}$Sn [Fig. \ref{width} (b)], whereas as a function of $M$ the width saturation takes place in $^{106}$Sn already at $T\geq$ 3 MeV [Fig. \ref{width} (d)]. In 
$^{88}$Mo, which is lighter than $^{106}$Sn, the GDR width keeps increasing   with $T$ up to $T\sim$ 5 -- 6 MeV at low $M$, but also starts to saturate at $M\geq$ 60 MeV [Figs. \ref{width} (a) and \ref{width} (c)]. At very high $M$ (above 100 $\hbar$) the GDR width ceases to change with $T$, but this
value of $M$ is too large to be realistic as will be discussed in the next section. The width saturation at high $T$ is a feature, which is obtained within the mechanism of GDR coupling to $pp$ ($hh$) configuration at $T\neq$ 0 and has been discussed previously (See Refs. \cite{PDM1,PDM2}, e.g.). The same mechanism takes place at $M\neq$ 0, where the contribution of GDR coupling to $pp$ ($hh$) to the width appears  whenever $f_{k'}^{\pm} - f_{k}^{\pm}$ is not zero [$(k,k') = (p,p')$ or $(h,h')$] [See Eq. (\ref{gamma1})]. At small $T$ the rotation frequency $\gamma$, which dictates the dependence the GDR width on $M$, clearly increases with $M$ at low $M$ but saturates at high $M$ as shown in Fig. \ref{ff} (a) for $^{106}$Sn at $T=$ 0.4 and 1 MeV.  The temperature dependence of the GDR width is determined essentially by that of the differences $f_{k'}^{\pm}-f_{k}^{\pm}$ at the right-hand side of Eqs. (\ref{gamma}) and (\ref{gamma1}). An example of such dependence is shown in Fig. \ref{ff} (b), where the factors  $f_{k'}^{\pm}-f_{k}^{\pm}$ for the pair of neutron levels $(1d_{5/2}-2p_{1/2})$, $(1d_{5/2}-2d_{5/2})$, and $(1g_{9/2}-2f_{7/2})$ in $^{106}$Sn,  calculated at $T=$ 4 MeV, are plotted as functions of $M$. It is clear from this figure that $f_{k'}^{\pm}-f_{k}^{\pm}$ at a given $T$ approach a saturation at high $M$. This means that, although at large $T$ the rotation frequency $\gamma$ increases with $M$ even at high $M$ [See Fig. \ref{ff} (a) at $T\geq$ 2 MeV], the width saturation caused by temperature seems to dominate, which leads to the combined effect of width saturation at high $T$ and $M$. 

Few experimental data, available only within narrow regions of $T$ and $M$, are also shown in Fig. \ref{width}. These are the GDR widths in $^{88}$Mo~\cite{Maj} and $^{106}$Sn~\cite{Mattiuzzi}, as well as the neighboring isotopes $^{109, 110}$Sn~\cite{Bra}, which are also spherical in the ground state (i.e. at $T=$0 and $M=0$).
Except for the data point in $^{88}$Mo at $T=$ 4 MeV and $M=$ 44 $\hbar$ [Figs. \ref{width} (a) and \ref{width} (c)], which will be discussed in the next section, the experimental data are found in fair agreement with the theoretical predictions.
%%%%%%%%%%%%%%%%%%%%%%%%%%%%%%%%%%%%%%%%
\subsection{Specific shear viscosity as a function of $T$ and $M$}
%%%%%%%%%%%%%%%%%%%%%%%%%%%%%%%%%%%%%%
\begin{figure}
     \includegraphics[width=16.0 cm]{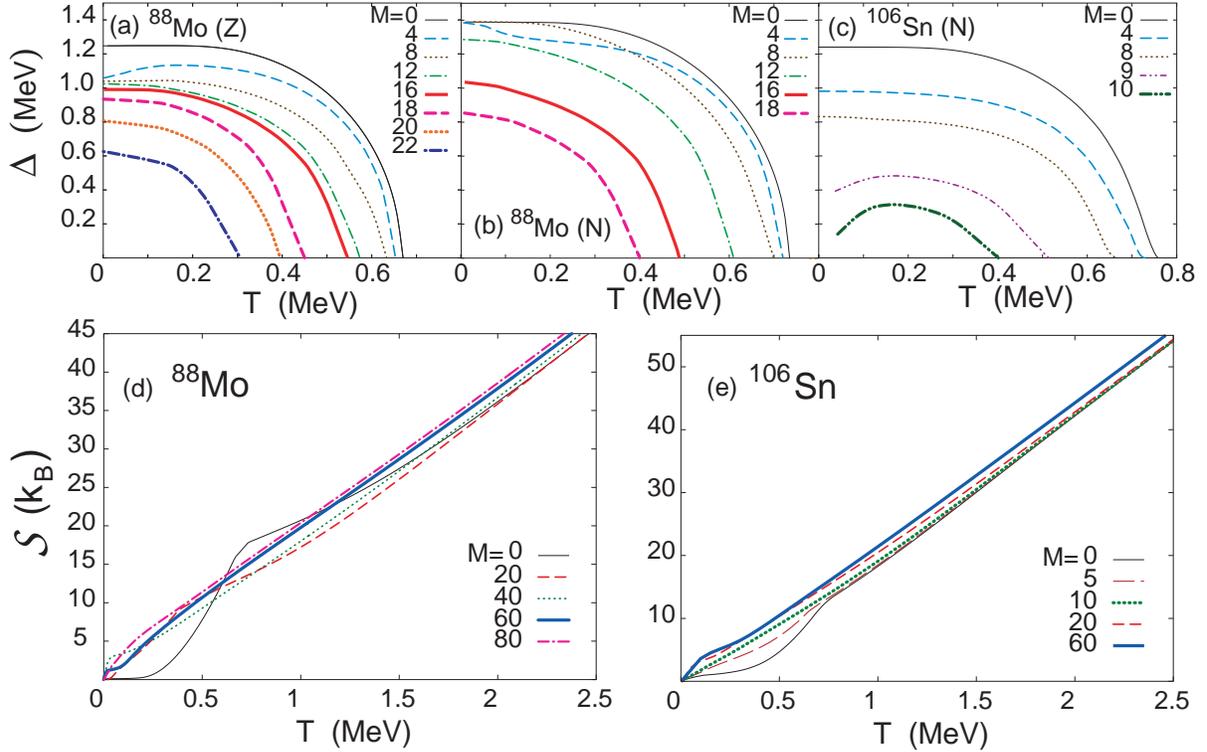}
     \caption{(Color online) (a) -- (c): Pairing gaps as functions of $T$ at different values of $M$ (in $\hbar$) as indicated in the panels for
     protons (a) and neutrons (b) in $^{88}$Mo and neutrons in $^{106}$Sn (c). Panels (d) and (e) show the entropies as functions of $T$ in $^{88}$Mo and $^{106}$Sn, respectively  at several values of $M$.}    
      \label{gapS}
\end{figure}
The BCS pairing gaps and entropies for $^{88}$Mo and $^{106}$Sn are shown in Fig. \ref{gapS} as functions of $T$ at several values of $M$. The general trend of the pairing gap is to decrease with increasing $T$ and $M$. But a slight pairing reentrance occurs for protons in $^{88}$Mo at $M=$ 
4 $\hbar$, and for neutrons in $^{106}$Sn at $M =$ 9 and 10 $\hbar$, where the gap increases with $T$ at $T<$ 0.2 MeV. For the detail discussion of this phenomenon, see Ref. \cite{SCQRPA} and references therein. The pairing gap vanishes at $M>$ 22, 18, and 10 $\hbar$ for protons, neutrons in $^{88}$Mo, and neutrons in $^{106}$Sn, respectively at all $T$. This feature can also be seen in the entropies, which become smoother as a function of $T$ with increasing $M$.
\begin{figure}
     \includegraphics[width=16.0 cm]{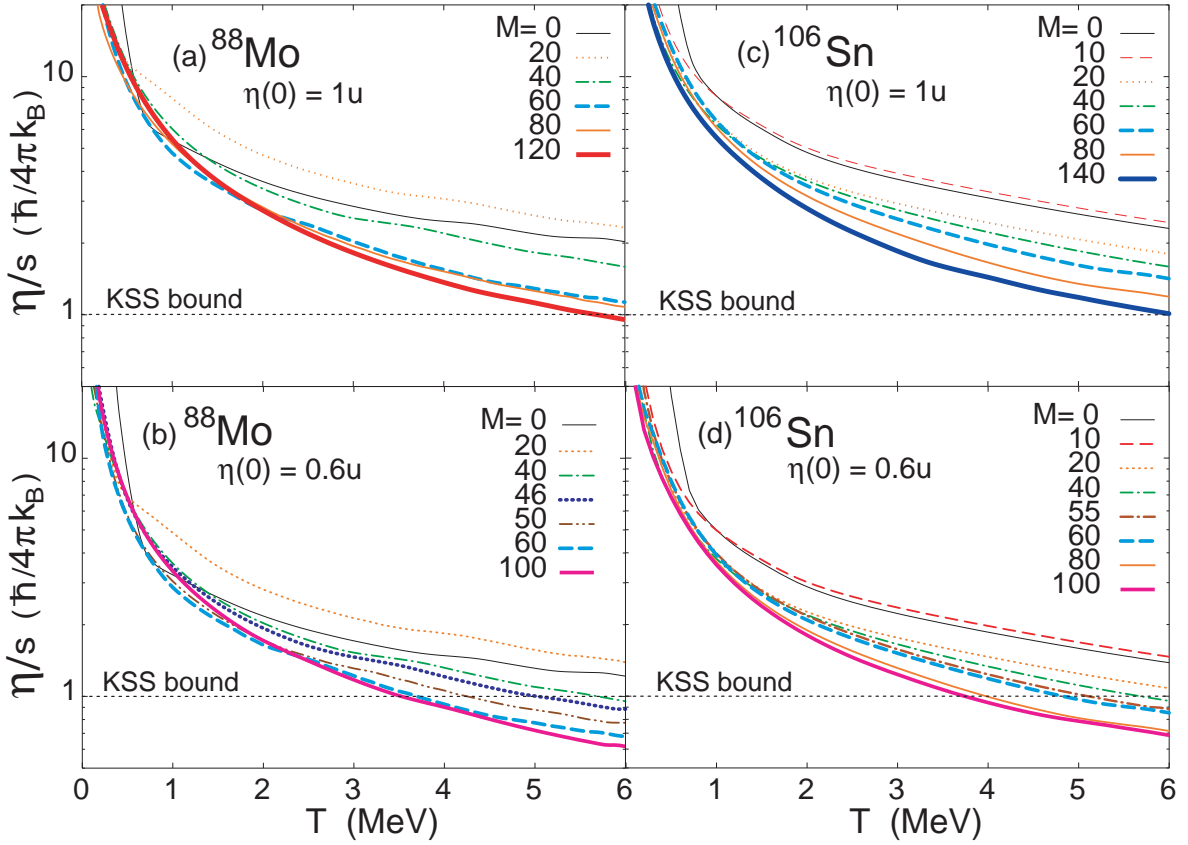}
     \caption{(Color online) Specific shear viscosity $\eta/s$ as a 
     function of $T$ at various $M$ (in $\hbar$) for $^{88}$Mo [(a) and (b)] and $^{106}$Sn [(c) and (d)]. 
     The results obtained by using $\eta(0) = 1~u$ are shown in (a) and (c), whereas those obtained by using $\eta(0) =0.6~u$ 
     are shown in (b) and (d) ($u = 10^{-23}$ Mev s fm$^{-3}$).}    
      \label{r}
\end{figure}
Shown in Fig. \ref{r} is the specific shear viscosity $\eta/s$ as a function of $T$ at several values of $M$ for $^{88}$Mo and $^{106}$Sn.
The results are obtained by using Eq. (\ref{eta}) at two values of $\eta(0)=$ 1$u$ [Figs. \ref{r} (a) and \ref{r} (c)] and 0.6$u$ 
[Figs. \ref{r} (b) and \ref{r} (d)]. The general trend of $\eta/s$ to decrease with increasing $T$ at a given value of $M$ is seen in all cases.
The ratio $\eta/s$ also decreases with increasing $M$ at a given $T$. By using $\eta(0)=$ 1$u$ it is seen that the value of $\eta/s$ is always larger than the KSS conjectured lower bound up to very high $T=$ 6 MeV and $M=$ 100 and 140 $\hbar$ for $^{88}$Mo and $^{106}$Sn, respectively. However, by using the lowest available value of $\eta(0)=$ 0.6$u$, extracted from the nuclear fission data, one finds the limiting values of $M$ to be
around 46 and 55 $\hbar$, respectively,  so that $\eta/s$ does not violate the lower bound conjecture up to $T=$ 5 MeV, which is the temperature where the GDR and a nucleus are assumed to cease to exist. These maximal values of $M$ are found in reasonable agreement with those obtained in the calculations of the critical angular momentum in the entrance reaction channel~\cite{Wilczynski} and the maximum angular momentum of the compound nucleus proposed in Ref. \cite{Grodzins}. 

\begin{figure}
     \includegraphics[width=14.0 cm]{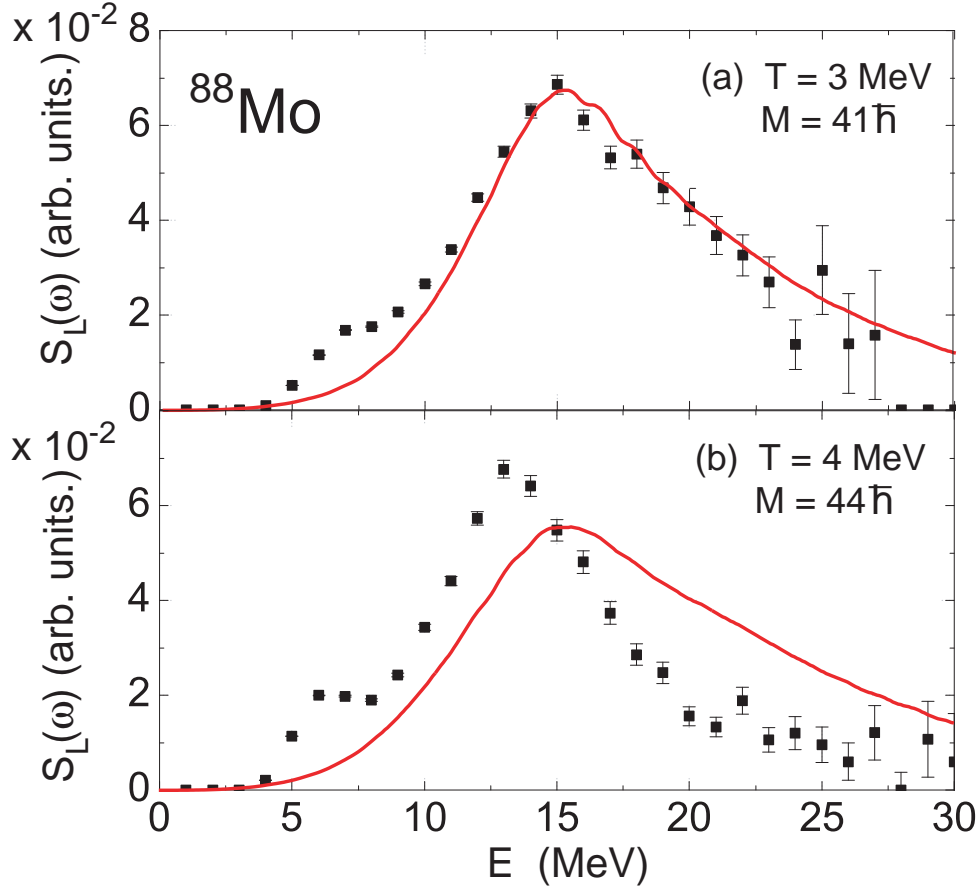}
     \caption{(Color online) GDR strength function $S_L(\omega)$ from Eq. (\ref{SL}) (in arbitrary units) for $^{88}$Mo at ($T=$ 3 MeV, $M=$ 41 $\hbar$) (a) and 
    ($T=$ 4 MeV, $M=$ 44 $\hbar$) (b) predicted by the PDM in comparison with the preliminary data from Ref. \cite{Maj}. }    
      \label{STS}
\end{figure}
Finally, by using the temperature dependence of $\eta/s$ and the KSS lower bound conjecture, it is possible to examine the reliability of the recent preliminary data for the GDR width in $^{88}$Mo reported in Ref. \cite{Maj}. 
As a matter of fact, the experimental GDR strength function is usually extracted from the GDR photoabsorption cross section, which can be well fitted with a Lorentz line shape. The latter is composed of two Breit-Wigner distributions [See Eq. (16) of Ref.~\cite{viscosity}, e.g.]. In the present case, such Lorentzian-like strength function reads 
 \begin{equation}
 S_L(\omega) = \frac{\omega}{E_{GDR}}[S(\omega, E_{GDR})-S(\omega, -E_{EGDR})]~,
 \label{SL}
 \end{equation}
 where $S(\omega,E_{GDR})$ is the strength function (\ref{SE}) with the main peak (or the location parameter of the distribution) located at $E_{GDR}$, whereas $S(\omega,-E_{GDR})$ is obtained from $S(\omega,E_{GDR})$ by replacing $E_{GDR}$ with $-E_{GDR}$.
 The PDM predictions of the GDR strength function $S_L(\omega)$ in $^{88}$Mo at ($T=$ 3 MeV, $M=$ 41 $\hbar$) and 
($T=$ 4 MeV, $M=$ 44 $\hbar$) are displayed in Fig. \ref{STS} in comparison with the preliminary data of Ref. \cite{Maj}. 
 The theoretical strength functions have been normalized so that the peak of the strength function, obtained at ($T=$ 3 MeV, $M=$ 41 $\hbar$), matches the maximum of the corresponding experimental GDR strength function.  The figure shows that, while the theoretical and experimental line shapes of the GDR agree fairly well at ($T=$ 3 MeV, $M=$ 41 $\hbar$) with the FWHM $\Gamma\simeq$ 11 MeV, they strongly mismatch at ($T=$ 4 MeV, $M=$ 44 $\hbar$), where the experimental GDR peak becomes noticeably narrower with a width $\Gamma_{ex}\simeq$ 7.5 MeV [See also Figs. \ref{width} (a) and \ref{width} (c)]. By using this value $\Gamma_{ex}$ and $\eta(0)=$ 0.6 $u$, one ends up with the value of $\eta/s=$ 0.85 KSS units. Including the error bars in $\Gamma_{ex}$ leads to $\Gamma_{ex}^{<}\simeq$ 6 MeV and $\Gamma_{ex}^{>}\simeq$ 8.5 MeV, which give the values of $\eta/s$ equal to 0.69 and 0.94 KSS, respectively. All these values are smaller than the KSS lower bound conjecture. Therefore, one can conclude that either (i) the data analysis of the GDR strength function
for $^{88}$Mo at $T=$ 4 MeV and $M=$ 44 $\hbar$ is inaccurate, or (ii) a violation of the KSS conjecture has been experimentally confirmed for the first time ever. A reanalysis of the data is expected to clarify which one from these two conclusions holds.
%%%%%%%%%%%%%%%%%%%%%%%%%%%%%%%%%%%%%%%
\section{Conclusions}
%%%%%%%%%%%%%%%%%%%%%%%%%%%%%%%%%%%%%%%%%%%%%%%%%%%%%%%
The present paper extends the PDM to include the effect of finite angular momentum on
the damping of GDR at finite temperature. The formalism is based on the description of the non-collective (single-particle) rotation
of spherical systems. This implies that the total angular momentum $J$ can be aligned along the $z$-axis, therefore it is completely determined by its projection $M$ on this axis alone.

The numerical calculations were carried out for two spherical nuclei $^{88}$Mo and $^{106}$Sn, for which the experimental data for the GDR width are available at $T\neq$ 0 as well as $M\neq$ 0 at sufficiently large values of $M$. The analysis of the numerical results show that the GDR width increases with $M$ at a given value $T$ for $T\leq$ 3 MeV. At higher $T$, the GDR width approaches a saturation at $M\geq$ 60 $\hbar$ for $^{88}$Mo and $\geq$ 80 $\hbar$ for $^{106}$Sn. However, the region of $M\geq$ 60 goes beyond the maximum value of $M$, up to which the specific shear viscosity $\eta/s$ has the values not smaller than the KSS lower bound conjecture for this quantity. This maximum value of $M$ is found to be equal to  46 and 55 $\hbar$ for $^{88}$Mo and $^{106}$Sn, respectively, if the value $\eta(0) =$ 0.6$\times 10^{-23}$ Mev s fm$^{-3}$ for the shear viscosity at $T=$ 0 is used.

A check by using the KSS lower bound conjecture for the specific shear viscosity and the same $\eta(0) =$ 0.6$\times 10^{-23}$ Mev s fm$^{-3}$ also shows that the experimental data for the GDR line shape in $^{88}$Mo at $T=$ 4 MeV and $M=$ 44 $\hbar$~\cite{Maj} leads to a violation of the KSS conjecture. This calls for the need of reanalyzing the recent experimental data reported in Ref. \cite{Maj} for the GDR in 
$^{88}$Mo  at these large values of temperature and angular momentum.
\acknowledgments
The numerical calculations were carried out using the FORTRAN IMSL
Library by Visual Numerics
on the RIKEN Integrated Cluster of Clusters (RICC) system. 

The author is grateful to A. Maj for fruitful discussions and M. Ciemala for help with the preliminary experimental data of GDR strength functions in $^{88}$Mo~\cite{Maj}. Thanks are also 
due to N. Quang Hung for assistance in numerical calculations.
%%%%%%%%%%%%%%%%%%%%%%%%%%%%%%%%%%%%%%%%%%%%

%%%%%%%%%%%%%%%%%%%%%%%%%%%%%%%%%%%%%%%%%%%%%%%%%
\end{document}